\documentclass[twoside,fleqn]{article}
\usepackage{espcrc2}
\usepackage{graphicx}

\title{
\vspace{-1.0cm}
{\sf \normalsize \rightline{HD-THEP-00-49}}
\bigskip
{\Large \bf QCD sum rule analysis of the charmonium system:
The charm quark mass}}
\author{M. Eidem\"uller\thanks{Talk given at the Euroconference on Quantum Chromodynamics
        (QCD 2000), Montpellier, July 2000} and M. Jamin
\address{
       {\em Institut f\"ur Theoretische Physik, Universit\"at Heidelberg,} \\
       {\em Philosophenweg 16, 69120 Heidelberg, Germany}} 
       }

\begin{document}

\newcommand{\nn}{\nonumber}
\newcommand{\mev}{\mbox{\rm MeV}}
\newcommand{\gev}{\mbox{\rm GeV}}
\newcommand{\eqn}[1]{(\ref{#1})}
\newcommand{\MSb}{{\overline{MS}}}
\newcommand{\ep}{\epsilon}
\newcommand{\IM}{\mbox{\rm Im}}

\begin{abstract}
In this work, the charm quark mass is obtained
from a QCD sum rule analysis of the charmonium system. 
In our investigation we include results from 
non-relativistic QCD at next-to-next-to-leading order.
Using the pole mass scheme, we obtain a value of
$M_c=1.70\pm 0.13$ GeV for the charm pole mass. The introduction
of a potential-subtracted mass leads to an improved scale
dependence. The running ${\rm \MSb}$-mass is then determined to be
$m_{c}(m_{c}) = 1.23 \pm 0.09$ GeV.
\end{abstract}

\maketitle

\newcommand{\jhep}[3]{{\it JHEP }{\bf #1} (#2) #3}
\newcommand{\nc}[3]{{\it Nuovo Cim. }{\bf #1} (#2) #3}
\newcommand{\npb}[3]{{\it Nucl. Phys. }{\bf B #1} (#2) #3}
\newcommand{\npps}[3]{{\it Nucl. Phys. }{\bf #1} {\it(Proc. Suppl.)} (#2) #3}
\newcommand{\plb}[3]{{\it Phys. Lett. }{\bf B #1} (#2) #3}
\newcommand{\pr}[3]{{\it Phys. Rev. }{\bf #1} (#2) #3}
\newcommand{\prd}[3]{{\it Phys. Rev. }{\bf D #1} (#2) #3}
\newcommand{\prl}[3]{{\it Phys. Rev. Lett. }{\bf #1} (#2) #3}
\newcommand{\prep}[3]{{\it Phys. Rep. }{\bf #1} (#2) #3}
\newcommand{\zpc}[3]{{\it Z. Physik }{\bf C #1} (#2) #3}
\newcommand{\sjnp}[3]{{\it Sov. J. Nucl. Phys. }{\bf #1} (#2) #3}
\newcommand{\jetp}[3]{{\it Sov. Phys. JETP }{\bf #1} (#2) #3}
\newcommand{\jetpl}[3]{{\it JETP Lett. }{\bf #1} (#2) #3}
\newcommand{\ijmpa}[3]{{\it Int. J. Mod. Phys. }{\bf A #1} (#2) #3}
\newcommand{\hepph}[1]{{\tt hep-ph/#1}} 
\newcommand{\hepth}[1]{{\tt hep-th/#1}} 
\newcommand{\heplat}[1]{{\tt hep-lat/#1}}



\section{Introduction}

An essential task within modern particle physics lies in the 
determination of quark masses which are key parameters
of the standard model.
In the past, QCD moment sum rule
analyses have been successfully applied for extracting the charm and
bottom quark masses from experimental data on the charmonium and 
bottonium systems respectively \cite{svz:79,rry:85,n:89}.
The fundamental quantity in these investigations is the 
vacuum polarisation function $\Pi(q^2)$:
\begin{eqnarray}
  \label{eq:a}
  \Pi_{\mu\nu}(q^2) &=& i \int d^4 x \ e^{iqx}\, \langle T\{j_\mu(x) j_\nu(0)\}\rangle \nn\\
  &=& (q_\mu q_\nu-g_{\mu\nu}q^2)\,\Pi(q^2),
\end{eqnarray}
where the current is represented by $j_\mu(x)=(\bar{c}\gamma_\mu c)(x)$.
Via the optical theorem, the experimental cross 
section $\sigma(e^+ e^- \to c\bar{c})$
is related to the imaginary part of $\Pi(s)$:
\begin{displaymath}
  \label{eq:b}
  R(s)=\frac{1}{Q_c^2}\,\frac{\sigma(e^+ e^- \to c\bar{c})}
  {\sigma(e^+ e^- \to \mu^+ \mu^-)}=12\pi\,\IM\, \Pi(s+i\ep).\nn\\
\end{displaymath}
Using a dispersion relation, we can express the moments by an integral
over the velocity $v=\sqrt{1-4m^2/s}$:
\begin{eqnarray}
  \label{eq:c}
 && {\cal M}_{n} = \frac{12\pi^2}{n!} \left(4m^2 \frac{d}{ds}\right)^n
  \Pi(s)\bigg|_{s=0} \\
 && \!\!\!\!\!\!\!\!\!\!\!\!\! =\! \left(4m^2\right)^n\!\! 
 \int_{s_{min}}^\infty \!\!\!\!\!ds\,\frac{R(s)}{s^{n+1}}
  = 2 \int_0^1 \!\!dv \, v(1-v^2)^{n-1}R(v).\nn
\end{eqnarray}
The moments can either be calculated theoretically, including 
perturbation theory, Coulomb resummation and non-perturbative contributions,
or can be obtained from experiment. In this way, we can relate the charm
quark mass to the hadronic properties of the charmonium system.

A natural choice for the mass appearing in eq. \eqn{eq:c} is the pole mass $M$.
In the first part of our numerical analysis, we will use the pole mass
scheme to extract the charm pole mass. However, as the pole mass suffers
from renormalon ambiguities \cite{b:99}, it can only be
determined up to corrections of order $\Lambda_{QCD}$. 
In the second part of
our analysis we shall therefore use the recently introduced 
potential-subtracted mass $m_{PS}$ \cite{b:98}.
From this mass definition we can obtain the ${\rm \MSb}$-mass
more accurately than from the pole mass scheme.

In the next section, we shall present the perturbative expansion 
of the correlator which is known up to next-to-next-to-leading order (NNLO).
As was found in recent analyses of the Upsilon-system 
\cite{kpp:98,my:98,pp:99,b:99:2,h:99},
the dominant theoretical contributions arise from the threshold expansion
in the framework of non-relativistic QCD (NRQCD). 
These contributions will be derived in
section 3. Afterwards, we shall shortly discuss the non-perturbative
contributions and the phenomenological spectral function.
In the numerical analysis, we shall first explain the method
of analysis.
We will then obtain the pole mass
and the ${\rm \MSb}$-mass from an analysis
in the pole mass and the PS-mass scheme respectively. 
The origin of different contributions to the error will be carefully
investigated. We shall conclude with a summary and an 
outlook.


\section{Perturbative expansion}

The perturbative spectral function $R(v)$ can be expanded in powers of the
strong coupling constant $\alpha_s$,
\begin{displaymath}
  \label{eq:d}
  R^{Pt}(s)= R^{(0)}(s)+ \frac{\alpha_s}{\pi}\,R^{(1)}(s)+ 
  \frac{\alpha_s^2}{\pi^2} \,R^{(2)}(s)+\ldots
\end{displaymath}
From this expression the corresponding moments
${\cal M}_n$ can be calculated via the integral of eq. \eqn{eq:c}.
The first two terms are known analytically and can for example be
found in ref. \cite{jp:97}. $\Pi^{(2)}(s)$ is still not fully known
analytically. However, the method of Pad{\'e}-approximants has been
exploited to calculate $\Pi^{(2)}$ numerically, using available
results at high energies, analytical results for the first eight
moments and the known threshold behaviour 
\cite{cks:96,cks:97}.
This information is sufficient to obtain a numerical
approximation of $\Pi^{(2)}(s)$ in the full energy range.

The numerical stability of the results can be checked in different 
ways. By choosing different Pad{\'e}-approximants or by selecting
a smaller set of input data the results for the moments remain
almost unchanged. Furthermore, some contributions to the spectral
density like those from internal quark loops are known analytically
and are in very good agreement with the numerical spectral density.
The above are strong indicators 
that the numerically obtained spectral density
comes very close to the exact spectral density.
  
 
\section{Coulomb resummation}

The perturbative expansion which contains terms up to the 
order $O(\alpha_s^2)$ is a good approximation for high velocities.
However, as one approaches the threshold region, terms of the order
$v(\alpha_s/v)^k$ become increasingly important. These terms can be obtained
in the framework of NRQCD. The correlator
is then expressed in terms of a Greens function \cite{pp:99}:
\begin{displaymath}
  \label{eq:e}
  \Pi(s)=\frac{N_c}{2M^2}\left(C_h(\alpha_s)G(k)+\frac{4k^2}{3M^2}G_C(k)\right),
\end{displaymath}
where $k=\sqrt{M^2-s/4}$ and $M$ represents the pole mass.
The constant $C_h(\alpha_s)$ is a perturbative coefficient which is
needed for the matching between the full and the non-relativistic 
theory. It naturally depends on the hard scale.
The Greens function is analytically known up to NNLO \cite{pp:99}
and sums up terms of order
$\alpha_s^n/v^{n-k}$ for $n\geq 0$ and $k=1,2,3$.
It is crucial for the analysis that the result depends on three scales.
While the hard scale $\mu_{hard}$ is responsible
for the hard perturbative processes,
the soft scale $\mu_{soft}$ governs the expansion of the
Greens function. Furthermore, the factorisation scale
$\mu_{fac}$ separates the contributions of large and small 
momenta and plays the role of an infrared cutoff.

The Greens function contains two parts: 
the continuum and the poles above and below threshold respectively.
We are interested in both contributions separately: first, the individual
corrections can be analysed and their error estimated. Second, in our
numerical analysis we will reconstruct the spectral density above 
threshold and we thus need the corresponding spectral density 
at low velocities.
In principle, the expressions for the energies and decay widths of the
poles have been calculated.
However, in the actual case of the charm quark this
expansion does not converge well. We will therefore choose a different
method of evaluation. Since the results for the Greens function are known
analytically, we can evaluate their contribution to the moments
numerically by performing the derivatives.
On the other hand, by using a dispersion relation, we can obtain
the continuum from the imaginary part of the correlator. 
From the difference we can obtain the pole contributions:
\begin{eqnarray}
  \label{eq:f}
  {\cal M}_n^{Poles} &=& \frac{12 \pi^2}{n!}\left(4M^2\frac{d}{ds}\right)^n
    \Pi(s)\Big|_{s=0} \nn\\
    && -12 \pi \left(4M^2\right)^n \int_{4M^2}^\infty ds\  
    \frac{\IM \,\Pi(s)}{s^{n+1}}.\nn
\end{eqnarray}
Since we will not evaluate the poles near threshold, but rather calculate
their contributions to the moments in a region where perturbation theory
is expected to be valid, the convergence of the pole contributions
is improved. Nevertheless, the poles will give the largest contribution
to the theoretical moments and thus the dependence on the
scales will remain relatively strong. 
The large corrections are partly due to the definition of the pole mass. 
These contributions can
be reduced by using an intermediate mass definition. In this analysis
we will use the potential-subtracted (PS) mass \cite{b:98} where
the potential below a separation scale $\mu_{sep}$ is subtracted:
\begin{eqnarray}
  \label{eq:f2}
  \delta m(\mu_{sep}) &=& -\frac{1}{2}\int\limits_{|{\bf q}|<\mu_{sep}}
  \!\!\!\frac{d^3 q}{(2\pi)^3}\,V(q),\nn\\
  m_{PS}(\mu_{sep}) &=& M-\delta  m(\mu_{sep}).\nn
\end{eqnarray}
As will be seen in the numerical analysis, this mass definition
leads to an improved scale dependence and a more precise
determination of the ${\rm \MSb}$-mass.


\section{Condensate contributions}

The non-perturbative effects on the vacuum correlator are parametrised
by the condensates. The leading correction is the gluon condensate
contribution which is known up to next-to-leading order \cite{bbifts:94}.
Furthermore, the dimension 6 and 8 contributions have been calculated
and will be included in our analysis \cite{nr:83:1,nr:83:2}. 
It will turn out that the condensate contributions are suppressed
when compared to former charmonium sum rule analyses and only have little influence
on the mass. Besides an increase of the theoretical moments from the
Coulomb contributions we
will restrict the moments to $n\leq 7$ where the non-perturbative contributions
are relatively small. Since we obtain a larger pole mass
than the former analyses, the condensates,
starting with a power of $1/M^4$, are suppressed further.


\section{Phenomenological spectral function}

Experimentally, the first six $\psi$-resonances have been observed.
Since the widths of the poles are very small compared to the masses,
the narrow-width approximation provides an excellent description
of these states. To model the contributions above the 6th resonance,
we use the assumption of quark-hadron-duality and integrate over the
perturbative spectral density:
\begin{displaymath}
  \label{eq:g}
  \frac{{\cal M}_{n}}{(4M^2)^n} = \frac{9\pi}{\alpha_{e.m.}^2 Q_c^2}
  \sum_{k=1}^6 \frac{\Gamma_k}{E^{2n+1}_k}
  + \int_{s_0}^\infty ds\,\frac{R^{Pt}(s)}{s^{n+1}}.
\end{displaymath}
To estimate the continuum contribution we will use a threshold
in the range of $3.6$ GeV $\leq \sqrt{s_0}\leq 4.2$ GeV with a central
value of $\sqrt{s_0}= 3.8$ GeV. However, the most dominant phenomenological
contributions come from the first two $\psi$-resonances resulting
in a small influence of the continuum even for low values of $n$.


\section{Numerical analysis}


\subsection{Analysing method}

Besides the contributions from the poles of the Greens function 
and the condensates, the theoretical part of the correlator
contains the spectral density above threshold. Now we will discuss
the different parts of the spectral density.
For high velocities the spectral density is well described
by the perturbative expansion. However, as one approaches smaller
values of $v$, the perturbative expansion breaks down. The resummed
spectral density, on the other hand, gives a good description
for low values of $v$, but becomes unreliable for high velocities.
For these reasons, we will introduce a separation velocity $v_{sep}\approx 0.3$.
Above $v_{sep}$ we will use the perturbative spectral density. Below
$v_{sep}$, we take the resummed spectral density adding the 
terms which are included in perturbation theory but not in resummation.

\begin{figure}
\includegraphics[width=7cm,height=4.5cm]{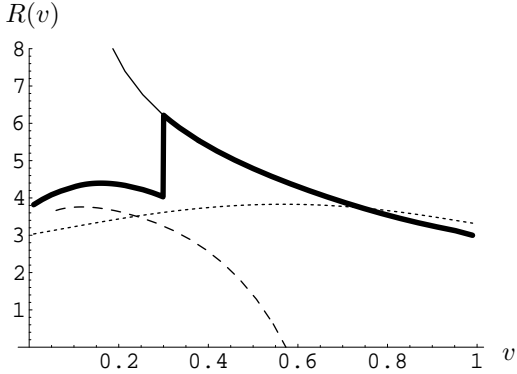}
\put(-6.7,4.7){$R(v)$}
\put(-.1,.2){$v$}
\caption{\label{fig:a}Thick solid line: reconstructed spectral density;
Thin solid line: perturbative spectral density;
dotted  line: perturbation theory at NLO;
dashed line: resummed spectral density.}
\end{figure}

In fig. \ref{fig:a} we have displayed the different contributions. The dotted
line represents the perturbative expansion in LO and NLO. The thin
solid line also includes the NNLO. Whereas for high velocities
the perturbative expansion is well convergent, the importance
of the higher corrections increases for smaller $v$. The dashed line
is the resummed spectral density and the thick solid line the 
reconstructed spectral density. For the charmonium system, there
exists a range of intermediate values of $v$ where neither the
perturbative expansion nor the resummation can be trusted. 
Indeed, it can be clearly seen that the reconstructed
spectral density shows a gap at the separation velocity.
To estimate the error we have varied $v_{sep}$ between 0.2 and 0.4.
The analysis shows that though the introduction of the separation
velocity stabilises the sum rules, the variation only has a minor 
influence on the mass.


\subsection{Pole mass scheme}

\begin{figure}
\includegraphics[width=7cm,height=4.5cm]{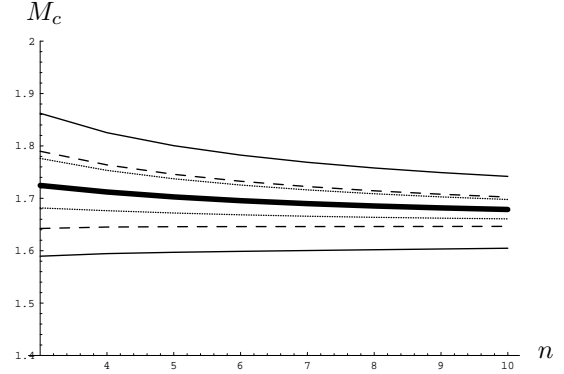}
\put(-6.7,4.7){$M_c$}
\put(0.1,.2){$n$}
\caption{\label{fig:b}
Thick solid line: central pole mass;
thin solid lines: $M_c$ for  $\mu_{soft}=1.05$
and 1.5 GeV;
dashed lines: $M_c$ for  $\mu_{fac}=1.2$
and 1.7 GeV;
dotted lines: $M_c$ for  $\mu_{hard}=1.4$
and 2.5 GeV.}
\end{figure}

Since the perturbative corrections grow for large values of $n$,
we will restrict our analysis to moments with $n\leq 7$. As our analysing
method needs moments of $n\geq 3$ to reconstruct the spectral density,
we will use $3\leq n\leq 7$. As the central values for our scales
we will choose 
\begin{eqnarray}
  \label{eq:h}
  \mu_{soft}&=& 1.2 \ \mbox{GeV}, \quad \mu_{fac}= 1.45\ \mbox{GeV},\nn\\
  \mu_{hard}&=& 1.7\ \mbox{GeV}. 
\end{eqnarray}
We have set the hard scale equal to the central value for the pole mass.
For the soft scale, we would have liked to choose a somewhat smaller
value, but the NNLO corrections get out of control for $\mu_{soft}\leq 1$ GeV.
The factorisation scale lies between the two other scales.
In this scheme the theoretical moments are dominated by the
pole contributions.
To estimate the error on
the mass, we have varied the scales within reasonable ranges.
The result is depicted in fig. \ref{fig:b}.
The thick solid line gives the central value for the charm mass
of $M_c=1.70$ GeV with the values of eq. \eqn{eq:h}.
The error is dominated by the variation of the scales, we obtain
\begin{eqnarray}
  \label{eq:i}
  1.05 \ \mbox{GeV} \leq \mu_{soft}\leq 1.5\ \mbox{GeV}:&&\!\!\!\!\!\!\!\! 
  \Delta M_c=100\ \mbox{MeV} \nn\\
  1.2 \ \mbox{GeV} \leq \mu_{fac\ }\leq 1.7\ \mbox{GeV}:&&\!\!\!\!\!\!\!\!
  \Delta M_c=50\ \mbox{MeV} \nn\\
  1.4 \ \mbox{GeV} \leq \mu_{hard}\leq 2.5\ \mbox{GeV}:&&\!\!\!\!\!\!\!\!
  \Delta M_c=40\ \mbox{MeV}. \nn
\end{eqnarray}
A significant uncertainty also comes from $\Lambda_{QCD}$.
\begin{table}[hbt]
\caption{\label{tab:a}Single contributions to the error of $M$.}
\begin{tabular}{lr}\hline
\multicolumn{1}{c}{Source} & \multicolumn{1}{c}{$\Delta M_c$} \\ \hline 
Variation of $\mu_{soft}$ & 100 MeV \\
Variation of $\mu_{fac}$ & 50 MeV \\
Variation of $\mu_{hard}$ & 40 MeV \\ 
Threshold $s_0$ & 10 MeV \\
Experimental error & 15 MeV \\
Variation of $v_{sep}$ & 10 MeV \\
Variation of $\Lambda_{QCD}$ & 50 MeV \\ \hline
Total error & 130 MeV \\ \hline 
\end{tabular}
\end{table}
By choosing $\Lambda_{QCD}=330\pm 30$ MeV we get an error of
$\Delta M_c=50$ MeV. The results are summarised in table \ref{tab:a}.
Adding the errors in quadrature we obtain the charm pole mass:
\begin{equation}
  \label{eq:j}
  M_c=1.70 \pm 0.13 \ \mbox{GeV}\,.
\end{equation}
Instead of deriving the ${\rm \MSb}$-mass from the pole mass, in
the following section we will
use the PS-mass to obtain a more precise value for the ${\rm \MSb}$-mass.


\subsection{Potential-subtracted mass scheme}

First, we have to choose a value for the separation scale $\mu_{sep}$.
This scale should be taken large enough to guarantee a perturbative
relation to the ${\rm \MSb}$-mass. On the other hand, it should be 
smaller than $M\,v$. Both conditions cannot be well fulfilled
at the same time. As a compromise value we will choose
$\mu_{sep}=1.0\pm0.2$ GeV. In this scheme the pole contributions
from the Greens function turn out to be smaller than in the 
pole mass scheme. The contributions from the condensates get more
important here and we shall restrict our analysis to $n\leq 6$ where
these corrections are under good control. 

\begin{figure}
\includegraphics[width=7cm,height=4.5cm]{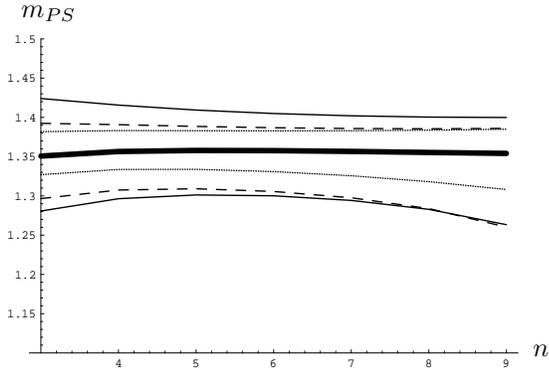}
\put(-6.7,4.7){$m_{PS}$}
\put(0.1,.2){$n$}
\caption{\label{fig:c}
Thick solid line: central PS-mass;
thin solid lines: $m_{PS}$ for  $\mu_{soft}=1.05$
and 1.5 GeV;
dashed lines: $m_{PS}$ for  $\mu_{fac}=1.2$
and 1.7 GeV;
dotted lines: $m_{PS}$ for  $\mu_{hard}=1.4$
and 2.5 GeV.}
\end{figure}
Using the same central 
values for the scales \eqn{eq:h} we obtain 
$m_{PS}(\mu_{sep}=1.0) = 1.35$ GeV and from this value a ${\rm \MSb}$-mass
of $m_{c}(m_{c}) = 1.23$ GeV. The introduction of the intermediate
mass definition leads to a reduced scale dependence:
\begin{eqnarray}
  \label{eq:k}
  1.05 \ \mbox{GeV} \leq \mu_{soft}\leq 1.5\ \mbox{GeV}:&&\!\!\!\!\!\!\!\!\!\! 
  \Delta m_{PS}=60\ \mbox{MeV}\nn\\
  1.2 \ \mbox{GeV} \leq \mu_{fac\ }\leq 1.7\ \mbox{GeV}:&&\!\!\!\!\!\!\!\!\!\!
  \Delta m_{PS}=40\ \mbox{MeV}\nn\\
  1.4 \ \mbox{GeV} \leq \mu_{hard}\leq 2.5\ \mbox{GeV}:&&\!\!\!\!\!\!\!\!\!\!
  \Delta m_{PS}=30\ \mbox{MeV}. \nn
\end{eqnarray}
\begin{table}[hbt]
\caption{\label{tab:b}Single contributions to the error of $m_{PS}$ and $m_{c}$.}
\begin{tabular}{lrr} \hline
\multicolumn{1}{c}{Source} & \multicolumn{1}{c}{$\Delta m_{PS}$} 
 & \multicolumn{1}{c}{$\Delta m_{c}$} 
\\ \hline 
Variation of $\mu_{soft}$ & 60 MeV & 50 MeV \\
Variation of $\mu_{fac}$ & 40 MeV  & 40 MeV\\
Variation of $\mu_{hard}$ & 30 MeV& 30 MeV \\ 
Variation of $\mu_{sep}$ & 30 MeV  & 30 MeV\\
Threshold $s_0$ & 30 MeV & 30 MeV\\
Experimental error & 10 MeV & 10 MeV\\
Condensates & 20 MeV & 20 MeV\\
Variation of $v_{sep}$ & 10 MeV & 10 MeV\\
Variation of $\Lambda_{QCD}$ & 10 MeV& 20 MeV \\ \hline
Total error & 90 MeV& 90 MeV \\ \hline 
\end{tabular}
\end{table}
In table \ref{tab:b} we have listed the individual contributions
to the error of $m_{PS}$ and $m_{c}$. Finally, we obtain for the masses
\begin{eqnarray}
  \label{eq:l}
  m_{PS}(\mu_{sep}=1.0) &=& 1.35 \pm 0.09\ \mbox{GeV}\nn\\
  m_{c}(m_{c}) &=& 1.23 \pm 0.09 \ \mbox{GeV}\,.
\end{eqnarray}
When we use the pole mass from eq. \eqn{eq:j} and calculate the
${\rm \MSb}$-mass using the three-loop relation between these masses 
\cite{mr:99,cs:99}, we obtain $m_{c}(m_{c})=1.20\pm 0.17$ GeV.
This value is in good agreement with eq. \eqn{eq:l}. This is not 
self-evident as the dominating pole contributions are reduced in the 
PS-scheme and the relative influence of the individual contributions
is shifted.


\section{Conclusions}

The obtained value for the pole mass lies somewhat higher
than in former sum rule analyses \cite{dgp:94,n:94:2}.
In  \cite{dgp:94} the authors used perturbation theory to NLO
resulting in a value of $M_c=1.46\pm 0.07$ GeV. In the second investigation
\cite{n:94:2} the analysis has been performed
in the ${\rm \MSb}$-scheme with perturbation theory to NLO. 
Using the NLO relation to the pole mass the author obtains
$m_{c}(m_{c}) = 1.26\pm 0.05$ GeV and $M_c = 1.42 \pm 0.03$ GeV.
The author has also performed an analysis using resummation
in LO with a value of $M_c = 1.45 \pm 0.07$ GeV.
In our analysis the increased value of the pole mass is essentially
due to large Coulomb contributions which have not been included in
former analyses. As a consequence, the error becomes larger as well. 
In a recent analysis, the pole mass has been estimated from the charmonium
ground state at NNLO \cite{py:98}. Here the authors obtained a pole
mass of $M_c = 1.88 ^{+0.22}_{-0.13}$ GeV. This value resulted from
large corrections of the Coulomb potential to the ground state.
During the last years, several lattice analyses obtained the following
values for the ${\rm \MSb}$-mass:
\begin{eqnarray}
  \label{eq:m}
   m_{c}(m_{c}) &=& 1.59 \pm 0.28\ \mbox{GeV}\ \  \mbox{\cite{accflm:94}},\nn\\
   m_{c}(m_{c}) &=& 1.33 \pm 0.08\ \mbox{GeV}\ \  \mbox{\cite{k:98}}, \nn\\
   m_{c}(m_{c}) &=& 1.73 \pm 0.26\ \mbox{GeV}\ \  \mbox{\cite{ggrt:99}}, \nn\\
   m_{c}(m_{c}) &=& 1.22 \pm 0.05\ \mbox{GeV}\ \ \mbox{\cite{bf:96}}.\nn
\end{eqnarray}
Whereas the results from \cite{k:98,bf:96} are in good agreement
with this analysis, the investigations from \cite{accflm:94} and \cite{ggrt:99}
obtain higher masses. For the time-being, the results are not conclusive
and future lattice calculations for the charm mass might be 
of interest.
Since other methods reveal significant uncertainties in the determination
of the quark masses, the sum rules remain one of the most precise
methods to extract these fundamental quantities.

\bigskip \noindent
{\bf Acknowledgements}
M.Eidem\"uller would like to thank S. Narison for the invitation to
this very ``charm''ing and interesting conference.


\begin{thebibliography}{99}

\bibitem{svz:79}
{\sc M.A. Shifman, A.I. Vainshtein and V.I. Zakharov},
\npb{147}{1979}{385}, \npb{147}{1979}{448}.

\bibitem{rry:85}
{\sc L.J. Reinders, H. Rubinstein and S. Yazaki},
\prep{127} {1985}{1}.

\bibitem{n:89}
{\sc S. Narison},
QCD Spectral Sum Rules, World Scientific (1989).

\bibitem{b:99}
{\sc M. Beneke},
\prep{317}{1999}{1}.

\bibitem{b:98}
{\sc M. Beneke},
\plb{434}{1998}{115}.

\bibitem{kpp:98}
{\sc J.H. K{\"u}hn, A.A. Penin and A.A. Pivovarov},
\npb{534}{1998}{356}.

\bibitem{my:98}
{\sc K. Melnikov and A. Yelkhovsky},
\prd{59}{1999}{114009}.

\bibitem{pp:99}
{\sc A.A. Penin and A.A. Pivovarov},
\npb{549}{1999}{217}.

\bibitem{b:99:2}
{\sc M. Beneke and A. Signer},
\plb{471}{1999}{233}.

\bibitem{h:99}
{\sc A.H. Hoang},
\prd{59}{1999}{014039}.

\bibitem{jp:97}
{\sc M. Jamin and A. Pich},
\npb{507}{1997}{334}.

\bibitem{cks:96}
{\sc K.G. Chetyrkin, J.H. K\"uhn and M. Steinhauser},
\npb{482}{1996}{213}.

\bibitem{cks:97}
{\sc K.G. Chetyrkin, J.H. K\"uhn and M. Steinhauser},
\npb{505}{1997}{40}.

\bibitem{bbifts:94}
{\sc D.J. Broadhurst et al.},
\plb{329}{1994}{103}.

\bibitem{nr:83:1}
{\sc S.N. Nikolaev and A.V. Radyushkin},
\npb{213}{1983}{285}.

\bibitem{nr:83:2}
{\sc S.N. Nikolaev and A.V. Radyushkin},
\plb{124}{1983}{243}.

\bibitem{mr:99}
{\sc K. Melnikov and T. van Ritbergen},
\plb{482}{2000}{99}.

\bibitem{cs:99}
{\sc K.G. Chetyrkin and M. Steinhauser},
\npb{573}{2000}{617}.

\bibitem{dgp:94}
{\sc C.A. Dominguez, G.R. Gluckman and N. Paver},
\plb{333}{1994}{184}.

\bibitem{n:94:2}
{\sc S. Narison},
\plb{341}{1994}{73}, \npps{74}{1999}{304}. 

\bibitem{py:98}
{\sc A. Pineda and F.J. Yndur\'{a}in},
\prd{58}{1998}{094022}.

\bibitem{accflm:94}
{\sc C.R. Allton et al.},
\npb{431}{1994}{667}.

\bibitem{k:98}
{\sc A. Kronfeld},
\npps{63}{1998}{311}.

\bibitem{ggrt:99}
{\sc V. Gim\'{e}nez et al.},
\npb{540}{1999}{472}.

\bibitem{bf:96}
{\sc A. Bochkarev and P. de Forcrand},
\npb{477}{1996}{489}, \npps{53}{1997}{305}.


\end{thebibliography}

\end{document}